# Transparent Programming of Heterogeneous Smartphones for Sensing

Felix Xiaozhu Lin[1], Zhen Wang[2], Robert LiKamWa[2], and Lin Zhong[1,2]
[1]Dept. of Computer Science and [2]Dept. of Electrical & Computer Engineering, Rice University

## Abstract

Sensing on smartphones is known to be power-hungry. It has been shown that this problem can be solved by adding an ultra low-power processor to execute simple, frequent sensor data processing. While very effective in saving energy, this resulting heterogeneous, distributed architecture poses a significant challenge to application development.

We present Reflex, a suite of runtime and compilation techniques to conceal the heterogeneous, distributed nature from developers. The Reflex automatically transforms the developer's code for distributed execution with the help of the Reflex runtime. To create a unified system illusion, Reflex features a novel software distributed shared memory (DSM) design that leverages the extreme architectural asymmetry between the low-power processor and the powerful central processor to achieve both energy efficiency and performance.

We report a complete realization of Reflex for heterogeneous smartphones with Maemo/Linux as the central kernel. Using a tri-processor hardware prototype and sensing applications reported in recent literature, we evaluate the Reflex realization for programming transparency, energy efficiency, and performance. We show that Reflex supports a programming style that is very close to contemporary smartphone programming. It allows existing sensing applications to be ported with minor source code changes. Reflex reduces the system power in sensing by up to 83%, and its runtime system only consumes 10% local memory on a typical ultra-low power processor.

## 1. Introduction

An emerging category of smartphone applications intend to serve their users in the background by sensing the physical world continuously, e.g., health monitoring, context-aware services, and participatory sensing. They, however, can significantly reduce the battery lifetime if left on. The key reason is that existing smartphones only expose the powerful central processor to general-purpose processing, including simple, frequent sensor data processing. Many researchers have shown that it is fundamentally necessary to add a low-power processor to process simple, frequent tasks, e.g., Turducken [3], Somniloquy [4], Little Rock [5], and Intel sensor hub [6]. The resulting system is *heterogeneous, distributed* with multiple processors loosely integrated and differing in their processing power by orders of magnitudes. For convenience' sake, we refer to the existing powerful central processor as the *central processor* and an added low-power processor as the *peripheral processor*.

The heterogeneous, distributed architecture, however, significantly challenges third-party application development, which has become the key to the success of commercial smartphones. Today's smartphone application developers only need to deal with the central processor (presumably a 32-bit ARM or X86 core); with the heterogeneous distributed architecture, they will have to deal with peripheral processors, which may have a very different microarchitecture, kernel, and programming language to be ultra-low power.

The goal of our work is to relieve smartphone app developers from the heterogeneous, distributed nature of the hardware so that they can easily leverage peripheral processors for sensing applications, without learning a new programming language or working with a new kernel.

While this goal is reminiscent of that of programming other heterogeneous, distributed systems, we face a set of unique challenges when dealing with smartphones. (*i*) We face an *extreme architectural asymmetry*. A peripheral processor shall be orders of magnitudes weaker than the central processor to tackle simple tasks with high efficiency. This renders many existing solutions targeted at high-performance systems inadequate, e.g., [7]. (*ii*) *Energy efficiency* is a dominant concern for smartphones while past work on heterogeneous distributed systems has focused on performance. As we will see later, this focus change will lead to very different designs in many aspects of the software system. (*iii*) Moreover, smartphones have already seen several *mature OSes and programming languages*. It is extremely difficult, if possible, to promote a clean-slate design of OS or programming model that could significantly reduce the system problems with heterogeneity, e.g., [7-9]. (*iv*) Finally, modern smartphones are increasingly multitasking. As applications are sharing the central processor today, they will share peripheral processors too. Therefore, existing programming solutions that dedicate a heterogeneous resource to a single task will not work.

Our solution to these challenges is a suite of runtime and compilation techniques, called *Reflex*. With Reflex, a developer only needs to encapsulate the candidate code to execute on a peripheral processor as a special class and limits access to system services in the code. Transformed by the Reflex compiler and assisted by the Reflex runtime, the encapsulated code will be able to run on a peripheral processor with native efficiency and communicate with the rest of the application as if it were also running on the central processor.

Reflex achieves the programming transparency not only by adapting known solutions in heterogeneous, distributed systems for the extreme asymmetric architecture, but also by innovative design and method. (*i*) The key to Reflex's transparency is a novel software distributed shared memory (DSM) design that allows shared memory objects between code running on multiple processors. Unlike existing software DSM solutions, the Reflex DSM achieves high energy efficiency without sacrificing too much performance. (*ii*) The Reflex runtime design judiciously exploits the computing power available on the central processor to reduce the hardware requirement to peripheral processors. It conceals the heterogeneous, distributed nature of the hardware from applications by using less than 6KB memory on a peripheral processor. (*iii*) The success of Reflex validates our key design principle in extremely asymmetric systems: heavily rely on the compile-time optimization to aggressively reduce the run time overhead.

Using a tri-processor heterogeneous smartphone prototype built with a Nokia N900 smartphone, a 32-bit microcontroller, and a 16-bit microcontroller, we report a complete implementation of Reflex for heterogeneous systems. We evaluate this Reflex realization using smartphone sensing applications reported in recent literature. By examining the source code of sensing applications implemented with and without Reflex, we show that Reflex only incurs less than 5% of source code difference, several lines of code in many cases. Experiments further demonstrate that Reflex significantly reducing the average power consumption in sensing (up to 83%). The evaluation in particular highlights the effectiveness of the Reflex DSM design, which not only maintains the high energy-efficiency of the heterogeneous architecture but also adds negligible execution time (<2.5%) to sensor data processing.

The rest of the paper is organized as follows. We motivate the importance of a heterogeneous, distributed architecture for smartphone sensing and outline the architectural assumptions made by Reflex in Section 2. We provide an overview of Reflex including its targeted transparency in Section 3. In Sections 4 and 5, we present the design of the Reflex runtime and DSM, respectively. We provide the prototype realization in Section 6 and report experimental results in Section 7. We discuss the related work in Section 8 and conclude in Section 9.

## 2. Heterogeneous Smartphones

We next provide the motivations to Reflex and outline the key assumptions made by the Reflex design.

### 2.1 Smartphone Sensing Applications

Smartphone-based sensing applications usually feature periodical tasks that access and process sensor data. The following observations motivate our work:

- Highly frequent tasks are usually very simple and can be adequately executed by a processor that is weaker than the central processor by orders of magnitudes.
- Events that require high computational power or Internet connectivity are less frequent than the simple, frequent tasks by orders of magnitudes.

Similar observations have been made and leveraged by researchers that build wireless sensor networks for event detection [10-13]. There are two important facts behind these two observations. First, events that are interesting to the applications must be rare. They must happen so rarely that processing them will not be a problem to the resource-constrained smartphone. Like in event-detecting wireless sensor networks, most resources will be consumed by simply detecting the presence of events, not by processing them.

Second, for computational efficiency, developers often optimize a sensing application so that the execution of a complicated task is conditional on the output of a simple task. For example, variants of decision tree classifier are widely used in efficient event/object detection and recognition so that most sensor data will be rejected as uninteresting without being fully processed [1, 2]. Similarly, a low-power sensor, e.g., accelerometer, can be used in the early detection of events so that a more expensive sensor, e.g., GPS receiver, is summoned only when an interesting event is highly likely [1, 14, 15]. According to our observation, most of the simple, frequent tasks of sensor data processing only require basic system support: computing, memory, sensor data, and timing.

### 2.2 Existing Heterogeneous Mobile Systems

It is well-known that a powerful processor is inefficient when it is lightly utilized, e.g., executing a simple task periodically [9]. Many architectural features intended for performance enhancement will not be fully utilized by a light workload, e.g., a deep pipeline, superscalar, speculative execution, and large cache. These features lead to high static power consumption and high overhead for power management, which inevitably makes a powerful processor inefficient for simple tasks. As a result, a homogenous system with a powerful processor such as that of today's smartphones is fundamentally inefficient for always sensing. For example, the accelerometer inside the N900 and iPhone 3GS consumes less than 1mW itself when active. Yet these smartphones will consume 50-100mW of power when simply reading the accelerometer at 30Hz, reducing their battery lifetime to barely acceptable.

A heterogeneous system is fundamentally necessary to realize sensing efficiently. Many have proposed heterogeneous architectures in various flavors for sensing applications on mobile devices, e.g., [3-6], and for wireless sensor nodes, e.g., [12, 13]. All these systems feature the use of two or more general-purpose processors that differ in processing power by orders of magnitude. All these systems employ a low-power "weak" processor to process the simple, frequent tasks and employ a more powerful processor to deal with demanding but rare tasks.

All above heterogeneous systems enjoy extreme asymmetry of orders of magnitudes in the computing power of the central and peripheral processors in terms of processing power and memory size, the power consumption in low-power modes, and the overhead for power management. We observe that this extreme asymmetry is the key to their effectiveness in conserving energy for always-on applications: the system spends a few milliwatts to stay alert with the weak processor and has the powerful strong processor available when interesting events happen.

## 2.3 Architectural Assumptions

We distill the following architectural features for future heterogeneous smartphone-based early prototypes [3-6] and our own experience in building heterogeneous smartphones. These features directly drive the Reflex design and are illustrated by Figure 1.

*Powerful Central Processor:* The central processor is likely to remain the most powerful engine for general purpose processing on smartphones. It is likely to become even more powerful for feature-rich applications, not only through a faster clock, but also more importantly through integrating more powerful cores. Reflex leverages the processing power of the central processor but seeks to keep it in a low-power state as much as possible.

*Multiple Microarchitectures:* A peripheral processor may have a different microarchitecture than the central processor. While weak ARM cores are available, 16-bit processors like TI MSP430 are necessary to achieve milliwatt power consumption for continuous sensing. This makes solutions based on single-ISA heterogeneous architectures inadequate.

*Multiple Kernels:* Instead of a monolithic OS, we adopt a multi-kernel approach as advocated by recent work for high-performance systems [7, 8]. The reasons are twofold. (*i*) Multitasking will be common for future smartphones and a local kernel allows localized resource management and higher efficiency. (*ii*) A multi-kernel solution can leverage existing kernels optimized for peripheral processors.

*Multiple Levels of Integration:* Peripheral processors are available through multiple levels of integration from various parties. Most reported heterogeneous mobile system prototypes are based on *board level* integration. Future system-on-a-chips (SoCs) for smartphones can integrate weak cores in the same chip, through either the system or peripheral bus. For example, TI OMAP4 SoC[16], intended for the next generation of smartphones, already integrates two low-power microcontroller cores on the system bus.

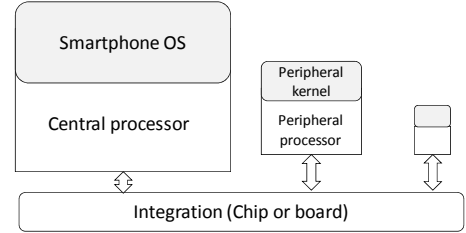

**Figure 1: The heterogeneous, distributed architecture of the smartphone platform intended for sensing, as envisioned by us and others.**

*Lack of Shared Memory:* Reflex does not assume a shared memory between the central and peripheral processors for the following reasons. (*i*) Accessing a shared physical memory is expensive for a weak processor: the memory controller can incur more power consumption than that by the processor itself by orders of magnitude (e.g. up to 80mW in TI OMAP3). (*ii*) As the processors may have different microarchitectures, shared memory hardware incurs further complication.

The architectural features described above immediately challenges third-party application developers who are used to program a single 32-bit ARM processor using a well-established programming style. None of existing heterogeneous mobile systems allow flexible and transparent programming. All except the mPlatform [9] and the Intel Sensor hub [6] requires the processors to be separately programmed, like separate systems. Therefore, the developers have to directly deal with the peripheral processor and its native kernel. To program the mPlatform as a single system, a special task-oriented abstraction must be used and tasks can automatically migrate between the processors. The Intel sensor hub allows only limited programmability of the weak processor through a predefined programming interface (API).

## 3. Reflex Overview

We first provide an overview of Reflex.

## 3.1 Targeted Programming Transparency

In the ideal case, a programmer should be completely unaware of the heterogeneous, distributed architecture. Yet, such complete transparency usually make system designs too sophisticated to be practical, which is known by the distributed system research [17] and also confirmed by our own investigation. Therefore, we relax the transparency goal of Reflex in two ways.

Reflex requires programmers to properly encapsulate candidate code to run on peripheral processors as *peripheral modules*. Accordingly, we call the code designated for the central processor the *central module*. A typical peripheral module retrieves the sensor data, processes it, and passes the processed results to the central module. By encapsulating simple, frequent tasks in a peripheral module executed on a peripheral processor, Reflex harvests the architectural heterogeneity for energy efficiency. This relaxation is reasonable because modern smartphone programming is already highly modular.

Reflex restricts system services that peripheral modules can use for two reasons: First, peripheral processors do not have all the resources available on the central processor, e.g., the file system and Internet connectivity; second, sensor data processing, according to our observation, only requires limited system services: dynamic memory, timer, and data acquisition.

### 3.2 Runtime and Compiler Support

Reflex achieves the targeted transparency with a distributed runtime and a compiler that transforms an application to execute with the runtime. With an application properly developed as discussed above, the Reflex compiler compiles the central module as a normal smartphone application and generates peripheral modules as in an intermediate representation (IR) format rather than architecture-specific binaries.

The runtime launches an application from the central processor, which maps modules to proper processors in the system. The central module starts and terminates as a normal smartphone process on the central processor, while the IR of peripheral modules are first translated into binaries for the peripheral processors the modules are mapped to, by invoking the corresponding static compilers. Then, the runtime ships the module binaries to proper peripheral processors for starting their execution. This mechanism effectively decouples a module from the processor it is actually allocated to, making such an allocation decision dynamic and open to the runtime. When the central module terminates and the application is going to end, the runtime will terminate all running peripheral modules in the application.

When two modules running on two processors exchange data through a procedure call or a shared memory object, Reflex automatically translates the procedure call into remote procedure call (RPC) and supports the shared object with software distributed shared memory (DSM). Both the RPC and DSM are implemented on top of the message layer of the Reflex runtime. The runtime and the DSM design will be presented in Sections 4 and 5, respectively.

### 3.3 Peripheral Module Abstraction

Reflex provides a virtual base class, called *ModuleBase*, to relieve developers from the lower layer programming details. ModuleBase declares a set of virtual methods as a concise interface for developers' code. Developers write a peripheral module as a class inheriting ModuleBase. A concrete module class consists of the developer's code for data processing (member methods) and states (member variables). For example, it overrides `OnCreate()` to initialize its states, and overrides `OnData()` to receive new sensor readings for processing. With the object-oriented style, developing peripheral modules is almost the same as the contemporary smartphone application development.

### 3.4 Hardware Requirement

Reflex requires a very small set of hardware features from peripheral processors and their integration with the central processor: at least 6KB ROM, several KB RAM, the capability to handle interrupts, and the capability to acquire sensor data without involving the central processor.

Without assuming physical shared memory among heterogeneous processors for reasons discussed in Section 2.3, Reflex only requires a simple data link for inter-processor communication. Such interconnect can be relatively low-speed (several hundred Kbps bandwidth), as long as it supports interrupt-driven, bi-directional message-passing, like I$^2$C.

Reflex does not require hardware Memory Management Unit (MMU). We are aware of that many existing distributed systems leverage hardware MMU to trap local memory operations in order to translate them as remote requests. However, it is very difficult, if not possible, to implement MMU for peripheral processors in Reflex, which is under so tight an energy constraint. This is partially confirmed by the absence of MMU in almost all commodity ultra-low power microprocessors. Therefore, Reflex employs the compiler to interpret proper memory access with inter-processor communication.

### 3.5 A Brief History of Reflex

Reflex has been an evolving system to provide system and programming support to an always-sensing mobile device with distributed, heterogeneous resources. The system is increasingly focused on facilitating third-party application development. The first generation of Reflex [18] defined a preliminary programming and execution unit, called *channel*, for code to be executed on microcontrollers. A daemon, precursor of the Reflex runtime, was employed to deploy channels and support channel-application communication. The developer must explicitly construct and pass messages for such communication.

Dandelion is a branch off the evolving Reflex system [19] to help smartphone application development with wireless body-area sensors. Because wireless body-area sensors are only loosely coupled with the smartphone sys-tem, we found RPC (RMI) adequate as the communication abstraction between code running on the wireless sensors and on the smartphone. The daemon of the first generation Reflex is replaced by a distributed runtime that implements message passing for RPC; the channel abstract is replaced by *senselet*, a platform-agnostic C++ class. Dandelion is able to conceal the heterogeneity of resources but its developers still have to cope with the distributed nature. For example, developers need to weave communication into computation.

The Reflex version reported here targets at completely hiding the distributed, heterogeneous nature from the developer with a unique DSM design. The introduction of DSM frees the developers from dealing with communication between distributed resources. At the same time, it not only introduced new Reflex components but also asked a completely new design of existing components of Reflex. (*i*) The Reflex compiler is new and plays critical role in the performance of DSM. (*ii*) The Dandelion senselet is replaced by Reflex *module*. While the Dandelion senselet is executed as a thread or process, the Reflex module embraces a specialized execution structure with a poll-based event queue at the core. This special structure is necessary to prevent DSM communication from overwhelming the peripheral processors. (*iii*) The message transport of the distributed runtime is redesigned to interface with the DSM and work with multiple processors. To support the DSM, the runtime now supports dynamic memory for message buffer management and monitors resources used in message passing and handling.

The current Reflex codebase contains more than 10K lines of code of which <10% is reused/modified from the Dandelion codebase. Both the peripheral runtime and compiler were developed almost from scratch.

## 4. Reflex Runtime System

The Reflex runtime has one component on each processor: the component on the central processor as the central coordinator, or *central runtime*, and other component runtimes running on peripheral processors are *peripheral runtimes*. Globally, component runtimes collectively form a lightweight message transport for communication among distributed modules. Locally, a component runtime provides modules with a unified abstraction of the resources and monitors resource usage. We leverage the processing power of the central processor in order to have a minimalist design for the peripheral runtime that can fit in ultra low-power processors.

### 4.1 Module Structure and Execution Model

Unlike many other heterogeneous systems that execute offloaded code synchronously, Reflex concurrently executes modules in native execution units as defined by the local kernel, e.g. thread or process. However, Reflex run-time defines modules with a specialized *event-driven* structure, different from generic threads or processes.

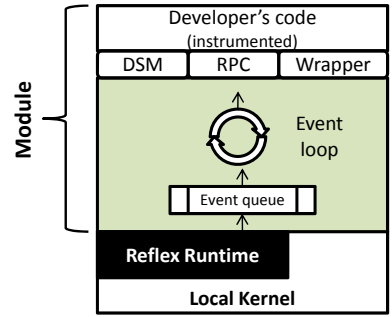

**Figure 2: The structure of a Reflex module. Arrows indicate the path of events in the module**

A Reflex event is a formatted data unit sent to a module from either the other modules or the local component runtime. These data units can be inter-processor communication messages, system services from the runtime (detailed in Section 4.3), or resource exception notifications (detailed in Section 4.4). A module has a private event queue to which the local component runtime asynchronously delivers events.

A module must *poll* its event queue to retrieve an event. Such a poll model provides the module and the compiler with the control over exactly when to handle asynchronous events. Polling occurs when the control flow returns from an event handler into the event loop. Polling also occurs according to the instrumentation by the Reflex compiler.

With the event-driven structure, a module consists of three code parts (as depicted in Figure 2): the event loop as the skeleton of the module, which keeps retrieving and dispatching events; the application-specific handling of various events, as provided by the developers, and functions that support DSM and RPC communications and wraps the system services. In a peripheral module, the event loop is realized in *ModuleBase*, and the application-specific event-handling is realized in a concrete module class.

The structure of the module has three major advantages. (*i*) Compared with a general procedural structure, the event-driven structure is widely adopted by contemporary smartphone applications, making the central module easily developed as a common smartphone application. (*ii*) Messages due to inter-module communication can be handled as a type of event as well, thereby simplifying the module skeleton code. (*iii*) Compared with the preemptive event handling, the polling model reduces data racing, since the execution of an event handler is atomic with regard to any other event, including messages sent from other modules of the application.

### 4.2 Message Transport

Built on top of the hardware interconnect, the runtime message transport serves as the lowest-level communication primitive among distributed modules. The endpoints of the transport are modules; and the basic unit of communication is a *message*. A message will be delivered as an event to the destination module. The destination module can later retrieve the message via polling.

Although in most aspects the message transport is a thin wrapper around the hardware interconnect, it supplies end-to-end delivery and thereby enables modules to communicate without knowing each other's physical location. To achieve this, each component runtime keeps a module location table, mapping all running modules to the processors they are running on. When a module is started or terminated, the central runtime sends the location update to all peripheral runtimes. This simple routing design is feasible since Reflex assumes a small number (<20) of processors and module location updates are rare.

The message transport is best-effort and does not guarantee reliable delivery. For example, a receiver module can silently drop a message if there is not enough free memory for buffering the message. Our rationales behind this decision are twofold. (*i*) The high-level protocols that the message transport supports, such as DSM and RPC, all use messages in request/response pairs between the two communicating modules. The response message implicitly indicates the delivery of the earlier request message. (*ii*) The best-effort choice reduces the communication and simplifies the message transport implementation by eliminating the delivery acknowledgement for each message.

### 4.3 Peripheral System Services

Reflex component runtime offers necessary local functionalities to support module execution. While the full-fledged central kernel already provides feature-rich support to the central module, the peripheral runtime provides only a compact set of peripheral system services, abstracting the resources of the underlying platform. The peripheral system services consist of a unified API available to all peripheral modules. *Sensor readings acquisition* allows a module to access raw sensor readings, either with the pull mode to acquire synchronously, or with the push mode to request delivering new readings as events. *Timer* is used to drive the data processing periodically by sending the module timer expiration events at the specified interval. *Dynamic memory* enables the module code to allocate memory space at execution time and free the space after use. New services that are critical to data processing can be added as they are supported by more heterogeneous units.

### 4.4 Monitor Resource Usage

The Reflex runtime monitors the usage of precious resources on peripheral processors by using peripheral runtimes as distributed 'detectors' that report to the central runtime. A peripheral runtime uses software counters to keep local statistics such as processor usage, dynamic memory, communication activities, etc. Local statistics can be queried by the central runtime as critical information to our long-term objectives in optimal allocation, energy accounting, and security.

The runtime system also watches for the emergent situations when resources run short or *resource exception* happens. Two resource exceptions can be raised by the peripheral runtime. *TimeException* is raised when the event queue of a module is close to overflow, indicating that the module lacks of processor time to handle events timely (e.g. processing sensor readings). *MemException* is raised when the stack of a module is close to overflow or dynamic memory allocation fails, indicating a lack of free memory space. Reflex exposes the exception handling to higher-level policies. For example, when TimeException is raised, the central runtime can choose to terminate the module launched most recently or the one consuming most processor time.

## 5. Distributed Shared Memory

The easiest way to exchange data between two software modules is through shared memory. Reflex provides a novel software distributed shared memory (DSM) because physical shared memory is highly unlikely between the central and peripheral processors. The Reflex DSM is highly energy-efficient, fits in the extremely asymmetric architecture, and is proved to be deadlock-free. It is completely transparent to programmers, as the Reflex compiler automatically translates memory objects shared by distributed modules into the DSM. There are numerous software DSM designs and realizations reported in the literature [20, 21] but none deals with the unique challenges as highlighted in *Introduction*. It is worth noting that even earliest software DSM systems back to late 1980s [22, 23] were built for machines with dedicated MMU support.

### 5.1 General Design Choices

Reflex implements DSM inside modules on top of the runtime message transport and relies on the compiler to instrument the developer's code. Compared to a runtime-level implementation, the module-level DSM grants the module code full control over how DSM operations are performed, and therefore enables the Reflex compiler to optimize the use of DSM operations on a per module basis.

*Memory Consistency Model:* In distributed systems, the memory consistency model defines the expected outcome from a series of memory operations from multiple processors. The Reflex DSM employs a variant of the release consistency [24] because a strict sequential consistency is energy-inefficient. The consistency achieved by the Reflex DSM is very close to that of traditional smartphone programming: for a given shared memory location, all writes are *sequential*; a read reflects a very recent write,

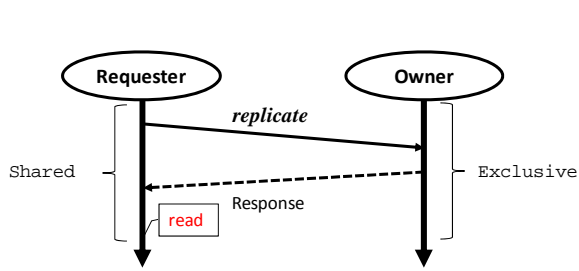
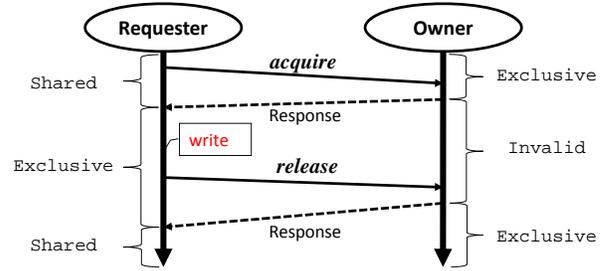

(a) *replicate* from a requester        (b) *acquire* and then *release* from a requester

**Figure 3: State transition in Reflex DSM coherence protocol is triggered by requests from a requester**

but not necessarily the most recent one. Such a timing gap in read is acceptable to sensing applications, as it is much smaller (by orders of magnitude) than the interval at which new processing results are produced.

*Coherence Granularity*: A DSM design must determine the basic memory unit for which coherence is maintained, i.e., the coherence granularity. Most DSM solutions leverage the MMU hardware and support the *page-level* granularity. That is, they employ a memory page, a continuous memory region with fixed size, as the basic memory unit for coherence. In contrast, the Reflex DSM is based on memory objects for two reasons. (*i*) Micro-architectural difference between processors makes the concept of memory page irrelevant: on different processors, same-size memory regions can hold different numbers of objects with distinct layouts. (*ii*) A fixed page size leads to a fixed tradeoff between stalls in execution and communication overhead of the resulting DSM, giving no opportunity to the compiler. In contrast, object-level granularity allows the Reflex DSM to aggressively leverage compile-time optimization to find out the best tradeoff.

## 5.2 Reflex DSM Coherence Protocol

We next present the *coherence protocol* based on the design choices made above. The protocol specifies how modules sharing an object should behave in accessing it.

### 5.2.1 Asymmetric Module Roles

Given a memory object in a smartphone application, we define its *sharing group* as all modules in the application that may access it. Each module in a sharing group has a local copy of the shared object. All write/read operations by a module on a shared object are actually performed on the local copy of the object.

In a sharing group, modules have asymmetric *roles* that are assigned by the Reflex compiler statically, according to the relative processing power of the processors that modules are intended to execute with. To facilitate the discussion, we mention a module as strong or weak in a group to refer to the relative processing power of its associated processor. The weakest module in the group is the *owner* of the object, while the other modules are *requesters* of the object. Since a module may access multiple shared objects, it can participate in multiple sharing groups and play different roles in them. Because the central processor is the most powerful processor in the system, the central module is always a requester.

Importantly, the owner behaves differently from a requester in accessing and synchronizing their local copies. In propagating the updated value of a shared object, a requester is *eager* and the owner is *lazy*. A requester sends the updated value to the owner actively; in contrast, an owner will send the updated value to a requester only when the requester asks for it. The owner's laziness guarantees the high energy-efficiency because DSM activities on a processor never wake up stronger processors. At the same time, the requester's eagerness ensures that the owner's stall period is small to avoid disrupting its data processing.

The role asymmetry employed by the Reflex DSM is different from existing DSM solutions in an important way. With performance as the primary goal, most existing DSM solutions employ the strong processor to host the owner role because the strong processor has rich resources to serve all requesters. In contrast, the Reflex DSM greatly favors energy-efficiency and, therefore, uses the weakest processor to host the owner role. Only if the weakest processor handles memory requests will the stronger processors be able to remain in sleep mode as much as possible.

### 5.2.2 Finite-state Machine Specification

We now provide details regarding the behaviors of the owner and requesters in terms of finite-state machines. A module in a sharing group can be in two possible states.

- The owner can be either `Exclusive` or `Invalid`.
- A requester can be either `Exclusive` or `Shared`.

In a sharing group, conceptually one and only one module is in the `Exclusive` state and it has the most updated value of the shared object in its local copy. A module in the `Exclusive` state can read and write to its local copy. The owner module in the `Invalid` state can neither read from nor write to its local copy; a requester module in the `Shared` state can read but not write to its local copy although the local copy may have an outdated value.

Modules in a sharing group change states only by requests sent by a requester. The owner module never in-itiates a change itself. A requester module uses three requests to change its own state and that of the owner module: *acquire*, *release*, and *replicate*.

Figure 3 (a) and (b) illustrate the interactions between a requester and the owner. In the two cases, the requester in the `Shared` state needs to read and write to the object, respectively. Before reading the shared object (subfigure (a)), the requester in the `Shared` state sends a *replicate* request to the owner. The owner responds by sending back the most updated value of the shared object. The *replicate* request will change neither the state of the requester nor that of the owner. Before writing to the shared object (subfigure (b)), the requester in the `Shared` state sends an *acquire* request to the owner in order to transit into the `Exclusive` state; it sends a *release* request (along with the latest value from the local copy) to the owner in order to transit back into the `Shared` state. The transition is performed when the requester receives the owner's response. Accordingly, the owner will transit to the `Invalid` and `Exclusive` states after responding to *acquire* and *release* requests, respectively.

In contrast to the active behavior of a requester, the owner module never actively seeks to change its own state. Before accessing the shared object, the owner in the `Invalid` state will stall its execution and passively wait until it is in the `Exclusive` state, usually after receiving a *release* request from a requester. The latency thereby introduced is called the *owner latency*.

The owner will respond to *acquire* and *replicate* requests only if it is in the `Exclusive` state. Otherwise, the requests will be queued and the owner will respond when it goes back to the `Exclusive` state. Before receiving the response from the owner, a requester will stall its execution. The latency thereby introduced is called the *requester latency*.

By keeping the owner in the `Exclusive` state as much as possible, latencies due to DSM will be minimized. The Reflex DSM achieves this with the following designs. First, a requester actively sends out a release request. Moreover, the owner lazily handles DSM requests. Other than the release request it is stalling for, the owner handles all DSM requests in its event loop. In the event loop, it retrieves a DSM request message only when there is no pending sensor data or timer event. This is because of the extreme asymmetry between the central and peripheral processors: requesters, running on stronger processors, can produce requests at a much higher rate than the owner, running on the weakest processor, can handle. This, of course, gives rise to the requester latency. However, we believe this price is worthwhile, since compared with the active data processing by the owner, requesters only occasionally use the shared object. Furthermore, the requester latency is mitigated by the compiler optimization described in Section 5.4.1.

## 5.3 Code Instrumentation by Compiler

The Reflex compiler adopts the following code analysis and instrumentation to realize the coherence protocol specified above.

The compiler performs alias analysis on the developer's code to statically infer the memory objects each program statement may read or write. With the analysis results, the compiler identifies shared objects and the corresponding sharing groups, in particular which module is the owner of a group.

For each access to a shared object by a module, the Reflex compiler inserts necessary code immediately before the access. The *pre-access* code must check the state of the module. It does nothing if the state is `Exclusive`. Otherwise, it calls DSM routines that are specific to the nature of the access (read or write) and the role of the module (owner or requester). Only after a write to the shared object in a requester module does the Reflex compiler insert *post-access* code to send out the release request and set the requester back to `Shared`.

Since the instrumented code will perform DSM operations according to the coherence protocol, overhead is introduced in both checking the module state and the intermodule communication (the state is not `Exclusive`). The Reflex compiler seeks to reduce the overhead through batching, described below.

## 5.4 Compiler Optimization with Batching

The Reflex compiler leverages two batching strategies in different ways for the owner and the requester modules. First, it is common that a shared object is repeatedly accessed in a small piece of the developer's code due to temporal locality, for example, in loop. By treating all these accesses as a single access and by only inserting the *pre-access* and *post-access* code before and after all of them, respectively, the Reflex compiler can suppress both the processing and communication overheads. This is called *intra-object batching*. Furthermore, it is also common that multiple shared objects are accessed in a small piece of developer's code due to spatial locality. By inserting the *pre-access* and *post-access* code for all the shared objects before and after all their accesses, respectively, the Reflex compiler can reduce the communication overhead because inter-module communication due to the shared objects can be aggregated. For example, multiple requests can be packed into a single Reflex message. This batching strategy is known as *inter-object batching*.

Also important to the performance of batching is the block size. A larger batching block tends to promote batching, thereby eliminating more DSM routine invocations within the batching block. However, it incurs two drawbacks as well. 1) Inter-object batching increases the chance of false sharing; because of more possible execution paths in a larger batching block, more acquired or replicated ob-jects may end up not being accessed. 2) Intra-object batching leads to longer stall of the object owner, since no *release* request is sent out within a larger batching block. On the contrary, a smaller batching block tends to suppress batching and eliminate fewer DSM routine invocations.

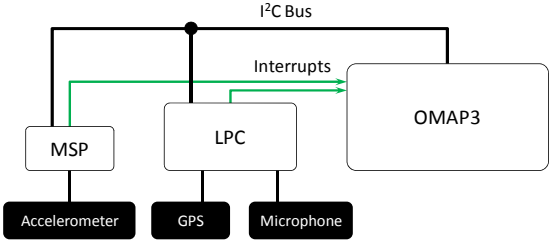

Figure 4: Architecture of the tri-processor hardware prototype

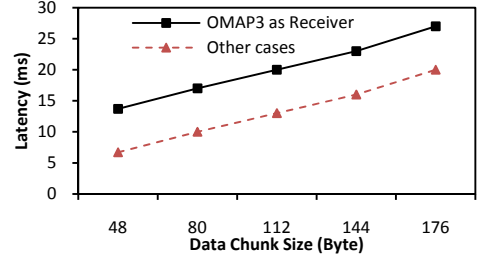

Figure 5: Latency in moving byte chunks on the I²C bus in the prototype. The chunk size is chosen as the typical Reflex message length

#### 5.4.1 Batching Strategy for the Requester Module

The Reflex compiler applies both inter-object and intra-object batching to the requester modules; using *program procedure* as a batching block. The primary performance concern of a requester is the latency in replicating/acquiring objects, which can be long due to the communication. Therefore, requesters benefit from the reductions in the number of requests, as enabled by inter-object and intra-object object batching. However, the batching block size is restricted by the requester's responsibility of timely releasing the shared object. Although moderately deferring release by requesters is tolerable to the owner (since the requesters are executed far faster than the owner), requesters still have to release the object before any blocking operation, e.g., sleep. As a result, the Reflex compiler uses the each program procedure as a batching block, which is large enough for effectively reducing requests and also ensures the acquired objects are released before calling any blocking procedure.

#### 5.4.2 Batching Strategy for the Owner Module

The Reflex compiler only applies intra-object batching to the owner module. Because the owner module never sends out any request, inter-object batching will not produce any benefit. For intra-object batching as an owner, the Reflex compiler starts with a *basic block* (code piece that is always executed sequentially) as the batching block and opportunistically extends the batching block to include multiple basic blocks within the same procedure. For each basic block, the Reflex compiler eliminates *pre-access* for a shared object if the object must have been accessed earlier in the same procedure. This is because earlier *pre-access* in the procedure already ensures the state is `Exclusive`. In executing a developer's procedure as a part of an event handler that is executed atomically, the state will never transit from `Exclusive` to `Invalid`.

Putting the two batching strategies together, the Reflex compiler instruments any application procedure in a module as follows: *i)* before each entry of the procedure, the compiler inserts the *pre-access* code for shared objects accessed by the procedure in the module role of requester; *ii)* before the entry of each basic block, the compiler inserts the *pre-access* code for shared objects accessed by this basic block in the module role of owner (subject to the intra-object batching); *iii)* after each exit of the procedure, the compiler inserts *post-access* for objects that have been acquired by the procedure in the module role of requester.

### 5.5 Deadlock Prevention

Within an application, deadlock may occur among multiple modules when one has already acquired some objects while trying to acquire more by waiting. Reflex DSM prevents the deadlock by applying the two regulations on the order of requests are sent and handled. First, in handling batched requests in a message, the owner module responds only when all requests in the message are satisfied with a single response message. Second, the Reflex compiler imposes a total ordering on all modules in an application, from the weakest to the strongest. In instrumenting the code of a requester, the compiler arranges all *pre-access* code according to the order of the object owners, and arranges all *post-access* code in the reverse order. With the two regulations, Reflex DSM is deadlock-free (see Appendix for a formal proof).

## 6. Prototype Realization

In order to evaluate the design, we have implemented Reflex based on a heterogeneous smartphone prototype.

### 6.1 Tri-Processor Hardware Prototype

We have built the envisioned heterogeneous architecture by extending a Nokia N900 smartphone. As shown in Figure 4, the prototype architecture employs N900's powerful OMAP3630 CPU (OMAP3) as the central processor and employs two ultra-low power microcontrollers, LPC1343 (LPC) and MSP430F1611 (MSP), as the peripheral processors. Table 1 summarizes the characteristics of these three processors. The prototype system uses Maemo Linux shipped with N900 as the central kernel and runs two separated μC/OS-II [25] on two peripheral processors. Peripheral modules are executed as μC/OS-II tasks.

*I²C-based board integration:* The three processors are integrated at the board-level via an I²C bus at 100KHz and share no physical memory. Since the Linux kernel of N900 does not support passive listening on I²C, the peripheral processors each use a GPIO pin to interrupt OMAP3. In order to physically access the I²C interface and GPIO lines of OMAP3, the N900's camera module is removed and its connector with OMAP3 is hijacked. Figure 5 shows the time spent in moving byte chunks on the I²C bus measured by a logic analyzer. The time serves as the baseline latency in inter-processor communication. As shown in Figure 5, a byte chunk with a size of a Reflex message usually takes tens of millisecond. When OMAP3 is the receiver, the measured time is on average 7ms longer due to the Linux context-switch latency after being interrupted by peripheral processors.

*Sensors:* N900, like all commercial mobile devices, tightly integrates its built-in sensors with the central processor (OMAP3). Therefore, it is very difficult for a peripheral processor to access the built-in sensors without waking up the central processor, violating the hardware requirement of Reflex as outlined in Section 3.4. A KXM52 tri-axis accelerometer is added to MSP through the local ADC interface; an analog microphone and an MN5010HS GPS receiver are added to LPC through the local ADC and the UART interface, respectively. This is also illustrated by Figure 4.

While the hardware prototype is built with off-the-shelf hardware components, it is neither the only nor the ideal incarnation of a heterogeneous smartphone. For example, one can optimize the peripheral processor design specifically for data processing to further improve the energy efficiency. We employ the architecture prototype mainly for the purpose of demonstration and evaluation of the ideas presented in Sections 3-5.

## 6.2 Reflex Prototype

We have implemented the complete Reflex design described in Section 4 for the tri-processor hardware prototype. It can work for all heterogeneous systems that meet the minimal hardware requirement summarized in Section 3.4 and have Maemo Linux as the central kernel.

### 6.2.1 Reflex Runtime

The Reflex runtime is implemented in 3800 lines of commented C/C++ code. The central runtime is a Linux system daemon that manages the execution of peripheral modules and talks to central modules over the dbus IPC facility. The central runtime only requires the I/O capability from the smartphone platform and, therefore, is highly portable. It can be easily ported to all smartphone platforms we have known. The peripheral runtime is implemented as platform-independent procedures added to the vanilla µC/OS-II kernel.

The message transport extensively uses dynamic memory as buffers to minimize its memory usage. This facilitates the zero-copy of messages: in delivering messages from the runtime to a module, only the pointer is deposited in the event queue and no message copy is needed. Overall, the transport incurs little computing overhead, which takes 1500 cycles (MSP), 891 cycles (LPC), and 1800 cycles (OMAP3) in sending a message, and 1560 cycles (MSP), 1612 cycles (LPC), and 1800 cycles (OMAP3) in receiving a message, excluding actual byte sending/receiving. Compared to the tens of milliseconds interconnect baseline latency (Figure 5), the computing overhead is negligible.

### 6.2.2 Module Library

**Table 1: Processors used in the tri-processor prototype**

|  | OMAP3 | LPC | MSP |
|---|---|---|---|
| **Clock Rate** | 600MHz | 72MHz | 3MHz |
| **Local Memory** | 256MB | 8KB | 10KB |
| **Active Power** | ~200mW | 42.9mW | 7.5mW |
| **Idle Power** | 13.4mW | 7mW | 3.2mW |

The module library implements the module structure and all Reflex routines that are executed as part of the module, including DSM and RPC. The library is linked with the developer's code in generating the final module binary. It has two versions, one for the central module and the other for the peripheral module. Both versions are implemented with 1000 lines of commented C/C++ code. Corresponding to the module structure described in Section 4.1, the library consists of three groups of routines:

- Realization of the event loop as the skeleton,
- Routines for RPC and DSM that can be invoked by the instrumented code, and
- Routines for wrapping the system services or IPC.

### 6.2.3 Reflex Compiler Toolkit

We build the Reflex compiler as a transformation pass on top of the LLVM compiler infrastructure [26] in 700 lines of commented C++ code. The Reflex compiler instruments the developer's code for DSM and compiles the code. The compilation output will be linked with the module library which is compiled by GCC.

The major task of the Reflex compiler is code instrumentation, which requires the knowledge about how shared objects are accessed in a given code piece. The Reflex compiler leverages the alias analysis framework provided by LLVM to infer the Mod/Ref information of a shared object, which indicate a program statement *must*, *never*, or *may* refer to, or modify the object.

For (un)marshaling objects across different ISAs, the Reflex compiler infers the data type of shared objects with best effort. The Reflex compiler statically analyzes the declarations of objects and tracks their use, to verify if their types are unambiguous. If the compiler is unable to determine the type of a shared object, it asks the developer for an explicit type description with IDL. We note that programmer-supplied type description is already a com-

```
Sensor Readings Acquisition:
Legacy:
  int ax, ay, az;
  FILE *f =
    fopen("/sys/class/i2c-3/3-001d/coord", "r");
  fscanf((FILE*) f, "%i %i %i", &ax, &ay, &az);
Reflex:
  int ax, ay, az;
  ax = GetSensorData(0);
  ay = GetSensorData(1);
  az = GetSensorData(2);

Timer:
Legacy: g_timeout_add(INTERVAL, OnTimer, TIMER_ID);
Reflex: RegisterTimer(INTERVAL, TIMER_ID);

Dynamic Memory:
Legacy: Mem = new int16_t [SIZE]; delete [] Mem;
Reflex:  Mem = (int16_t *)mod_malloc(SIZE); mod_free(Mem);
```

**Figure 6: System service invocations, as the only major difference in sensor data processing between Legacy and Reflex implementations**

**Table 2: Source lines of code of benchmarks**

|  | Reflex | Legacy |
|---|---|---|
| **Pedometer** | 199 | 200 |
| **uWave** | 220 | 215 |
| **RAPS** | 303 | 301 |
| **SoundSense** | 186 | 185 |

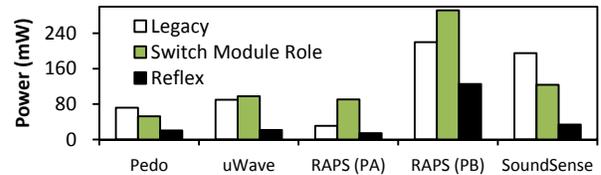

**Figure 7: The system power consumption in application scenarios of sensor data processing**

mon practice in smartphone development (e.g. AIDL in Android and dbus in Maemo).

The Reflex compiler forbids shared objects that embed pointer values. Lacking of a global address space, pointer values become irrelevant across distributed modules. While it is feasible to make pointers "portable," e.g., see [27], the Reflex prototype simply forbids sharing pointers across modules by rejecting any shared object that contains pointer and any type cast that is between pointer and integer in the developer's code (unless the cast is marked as trusted). In developing benchmark applications with Reflex, we found the above constraints reasonable and did not see any restriction from them on the development.

Reflex includes a small stub compiler that automatically produces data format conversion code for objects for inter-module data exchange. Implemented with around 500 lines of Python code, the stub compiler takes the IDL description (either generated by the Reflex compiler or supplied by the developer) to produce (un)marshaling code that is used for RPC and DSM.

## 7. Evaluation

We evaluate Reflex with real-world sensing applications and the tri-processor prototype reported in Section 6.

### 7.1 Benchmark Applications

Since there is no standard sensing application benchmark, we employ a combination of classic and most recently reported applications as benchmarks. We implement applications on N900's programming framework (Legacy) and on a tri-processor Reflex prototype (Reflex). The source lines of code are shown in Table 2. As noted in Section 6.1, Legacy implementations use the N900's built-in sensors while Reflex implementations use sensors connected to the peripheral processors.

Pedometer (*Pedo*) uses accelerometer readings to count the user steps. *Pedometer* (Reflex) consists of a peripheral module that estimates and counts steps; and a central module that provides a user interface (UI) to query the step count. The two modules share four integers that store step count, stride, velocity, and distance. The peripheral module can be executed by the MSP430.

*uWave* recognizes user gesture with the accelerometer and the open-source uWave algorithm [28]. When an interesting gesture is recognized, the smartphone UI is activated. *uWave* (Reflex) consists of two modules: a peripheral module periodically performs acceleration pattern matching and calls the central module when a gesture is recognized. The two modules share two objects storing the acceleration templates, each of which is a 64-element integer array. The peripheral module can be executed by the MSP430.

Rate-Adaptive Position Sensing (*RAPS*) [1] calculates geo-location uncertainty by comparing current acceleration activity with context-specific historical averages of velocity and acceleration activity. When the uncertainty passes a threshold, the application takes a GPS reading and updates the historical measures. *RAPS* (Reflex) consists of two peripheral modules PA and PB, and the central module. PA monitors acceleration and calculates location uncertainty. When the uncertainty exceeds a threshold, PA calls PB. PB then gathers GPS data, updates the recorded geo-locations, and provides PA with the relevant historical averages. The central module provides the UI to query the recorded geo-locations. PA and PB share the relevant historical averages. PB and the central module share all recorded timestamped geo-locations. PA and PB can be executed by the MSP and LPC, respectively.

*SoundSense*, adapted from [2], recognizes mobile user context by periodically analyzing microphone samples. If the energy or entropy of the frame passes a threshold, *SoundSense* captures a 5s sound window to determine the mobile user context. *SoundSense* (Reflex) uses a peripheral module to run the frame admission algorithm and a central module to execute context classification. The admission parameters (energy and entropy threshold) and the admitted window features are shared between the two modules. The peripheral module can be executed by the LPC.

Table 3: The percentage of estimated DSM latency in the execution of sensor data processing

|  | State Check | Owner Stall | Assumption |
|---|---|---|---|
| *Pedo* | 0.007% | 0.063% | Step detected every 400ms, UI query every 30 sec |
| *uWave* | 0.05% | 0.002% | UI activated every 15 min |
| *RAPS* | 0.001% | 0.008% | 350 new geo-locations per day, GPS on 27% of time [1] |
| *SoundSense* | 0.02% | 2.5% | Average frame admission rate 1.33Hz [2] |

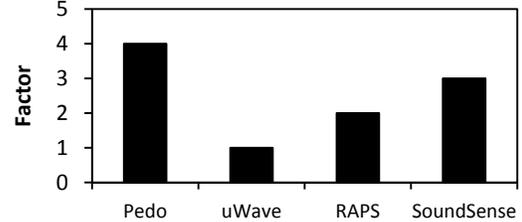

Figure 8: Factor of requester latency without inter-object batching to requester latency with inter-object batching

Due to their cascading staged structures and the sparsity of interesting sensory information, these applications spend most of their execution time in common scenarios of simple sensor data processing. These common scenarios determine the energy characteristics of the application.

### 7.2 Source Code Examination

We study the efforts in developing sensing applications by examining the source code of the benchmarks. The examination shows that Reflex effectively provides the targeted transparency for programming a heterogeneous smartphone. For the same benchmark, the Reflex and the Legacy implementations require very similar development efforts reflected in the source lines of code (Table 2) and the fact that they share most of the source code (95% identical source lines). Their source code only differs in the way of gluing the programming framework and invoking the system services described in Section 4.3 (code shown in Figure 6). This observation validates that Reflex has achieved its goal of the programming transparency by maintaining the contemporary programming style. Moreover, Reflex facilitates porting legacy sensing applications to the tri-processor prototype with only minor changes in the source code.

### 7.3 System Energy Efficiency

We measured the system power consumption of the prototype at 100Hz with the USB-2533 DAQ from Measurement Computing. For Legacy implementations, the system power only includes that by N900. For Reflex implementations, the system power includes all added hardware of the tri-processor prototypes. However, unused hardware is powered off in each particular benchmark.

Figure 7 shows the measured system power consumption for the most common scenarios of each benchmark. Not surprisingly, in scenarios where only the peripheral processors are executing data processing, Reflex implementations lead to significant power reduction, up to 83%. Only in scenarios when the central processor is involved do Reflex implementations carry the possibility of incurring power overhead (~3%) due to the added energy consumption of peripheral processors. As a result, in real life usage where the former scenarios dominate, Reflex implementations will be much more energy-efficient than legacy implementations. For example, *RAPS* (Reflex) will lead to 46% overall power reduction, estimated from the reported field trial [1].

The module role assignment in Reflex DSM (as described in Section 5.2) is the key to the energy efficiency goal. We have experimentally verified that assigning opposite module roles (the stronger processor as the object owner) will severely degrade the energy efficiency. This is due to the central processor being frequently wakened by *acquire* or *replicate* requests from the peripheral processors. As shown in Figure 7 ('switch module role'), the system power consumption will increase by up to 535%.

### 7.4 DSM Performance

Overall, DSM incurs very little overhead. In all four benchmark applications, we estimate that DSM routines only take at most 2.5% of the execution time in sensor data processing (as shown in Table 3).

In the common application scenarios where no communication occurs, the DSM overhead only comes from checking the module state. As we discussed before in Section 5.3, each state check only consists of several instructions. In Table 3, we show the estimated percentage of state check time in executing the sensor data processing. The results show that the overhead of state check is negligible (0.05%) even with *uWave*, which has the most frequent access of shared objects.

DSM introduces the requester latency and the owner latency only when multiple modules are accessing the same shared object. Fortunately, this only affects a very small portion of execution time. In the benchmarks, the requester latency is 41ms on average, which is acceptable to the UI interaction and the occasional object access by all central modules. The requester latency mainly consists of the time spent in the message transport (26ms~33ms) and in the owner lazy handling (~10ms). The owner latency is 20ms on average, dominated by the message transport time. Table 3 also shows the estimated percentage of the owner latency in executing the sensor data processing, up to 2.5% (*SoundSense*).

The batching strategy effectively amortizes the overhead over multiple objects and multiple accesses. We experimentally validate this by disabling a certain type of batching in the code instrumentation. The results in Figure 8 show that without inter-object batching, the overall requester latency can increase by up to a factor of 3. Intra-object batching particularly helps when shared objects are accessed in loops, e.g. the uWave matching algorithm iterates on the template with a two-level nested loop. The experiment also shows that without intra-object batching, uWave spends 1350 times of processor cycles on the state check in *pre-access*, which incurs 30% extra execution time in data processing.

## 7.5 Reflex Memory Overhead

Reflex introduces small memory overhead, especially for the peripheral processors with limited local memory. The introduced memory overhead has three parts: the runtime system, the module library, and the instrumented DSM code in the module. All three are small. (*i*) A peripheral runtime only takes a small portion of memory, 9.6% ROM on MSP and 16% ROM on LPC as shown in Table 4. (*ii*) The module library in each peripheral module takes 3% of total ROM on MSP and 5.4% on LPC. It is worth noting that since MSP is the weakest processor in the system, modules running on it always own shared objects. Therefore, the modules only link the DSM owner functions in the library. (*iii*) The instrumented DSM code only adds limited code to modules. For the two peripheral processors, each *pre-access* or *post-access* only adds around 10 bytes to the binary code. Table 5 shows that the instrumented *pre-access* and *post-access* code only increases the peripheral module binary size by at most 3%. In each central module, the instrumented code ranges from 200 to 700 bytes, less than 1.2% of the binaries.

## 8. Related work

Reflex is the first work that seeks to achieve transparent programming of smartphones with a heterogeneous, distributed architecture. The problem of programming heterogeneous, distributed systems is, however, not new. We next discuss the major approaches using representative solutions. Note some systems employ more than one approach.

*Directly programming*: Many heterogeneous distributed systems, including most existing heterogeneous mobile systems e.g. [3, 4] require developers to program each processor in the system directly. Code on two processors can only communicate through a narrow interface, often as two independent systems. The lack of programming trans-parency prevents this approach from being widely adopted by third-party developers.

**Table 4: Breakdown of Reflex memory overhead, in the format of ROM/RAM, in byte**

|  | MSP | LPC | OMAP3 |
|---|---|---|---|
| **Runtime** | 5446/82 | 5204/82 | 35748/647 |
| **Module Library** | 172/0 | 1546/0 | 11724/126 |

**Table 5: The bytes of instrumented code in peripheral module binaries**

|  | Instrumented Code | Increase % |
|---|---|---|
| **Pedometer** | 72 | 3% |
| **uWave** | 36 | 2% |
| **RAPS (PA)** | 72 | 3% |
| **RAPS (PB)** | 62 | 3% |
| **SoundSense** | 36 | 2% |

*Accelerator procedure*: Many systems offload computing tasks to heterogeneous hardware for synchronous, accelerated execution. This is particularly the case for recent work in GPGPU [29]. This approach executes the computing task as a procedure call to the heterogeneous hardware, synchronously with the application body. Data exchange between the application body and the offloaded code is relatively simple: through procedure arguments and the return value. However, this approach is limited in programming flexibility and does not support for multitasking.

*VM-based execution environment*: To make the system heterogeneity complete invisible to developers, an unified execution environment, mostly based on virtual-machine approach can be used to hide ISA difference. The virtual machine, however, is prohibitively resource-intensive for a heterogeneous smartphone with very weak peripheral processors. For example, Helios [7] requires processors of 32MB RAM and a few hundred MHz, much higher than that most peripheral processors can afford.

*Domain-specific programming model*: Custom programming models have been studied for heterogeneous systems. They include the stream-based programming in active disk for hard disks [30], capsule in active network for routers [31], and asynchronous tasks in CoMOS [32] for wireless sensor nodes. Although they are successful in their targeted application domains, they do not support the legacy programming model already used by third-party smartphone developers.

*Software DSM*: Many software DSM have been implemented, targeting performance using the compiler techniques [33] and runtime support [20, 23, 34]. Also, heterogeneity has been addressed in software DSM systems [35]. However, energy efficiency, as the dominant design goal of the Reflex DSM, leads us to a completely different design and distinguishes the Reflex DSM from all existing ones.

# 9. Conclusion

Sensing applications are considered to be one of the future killer applications for smartphones. While much research has been devoted to novel application development, its algorithmic foundation, and even heterogeneous system architecture, Reflex is the first endeavor toward making programming sensing applications for heterogeneous smartphones easier.

Despite decades' effort in programming heterogeneous, distributed systems, heterogeneous smartphones pose unique challenges that were previously not important. In addressing these challenges, we have not only revised known solutions but also devised novel ones in software DSM and distributed runtime. In particular, while the extreme architectural asymmetry of heterogeneous smartphones challenges existing solutions for programming heterogeneous, distributed systems, the success of Reflex highlights that the asymmetry can actually be exploited.

The success of Reflex in achieving its design goal is also a strong indication of the power of compile-time optimization. The extensive code instrumentation eliminates the requirement of MMU, making Reflex DSM fit in ultra low-power processors. Also, the batching technique aggressively optimizes the DSM performance.

We note that Reflex does not achieve complete transparency in programming due to the requirement to encapsulate a peripheral module and the limit on the peripheral system services. However, our experience suggests such relaxation is necessary and extremely profitable for sensing applications on smartphones. A complete transparency would be too expensive to be practical, especially with such an asymmetric architecture.

# Appendix: Deadlock Free Property

We next provide a formal proof that Reflex DSM is deadlock-free, as mentioned in Section 5.5. Our proof strategy is to assume a general deadlock situation and then show it is impossible.

As shown in Figure 9, a set of modules are in deadlock. They each hold exclusive access to some memory objects while waiting for exclusive access to more objects, thereby *circularly waiting* for each other, a necessary condition for deadlock [36]. In the figure, each box represents one or more objects. Objects in the same box, referred as an object set, must be owned by the same module, since requests to the same owner are handled atomically. We use Owner($n$) to refer the module that owns object set $n$.

We use A>B to denote that module A is stronger than module B, and A=B to denote that A and B are actually the same module.

Let's zoom in any single module, e.g. A, in the deadlock circle. In the example, A has already got the exclusive access to object set 1 and it is waiting for exclusive access

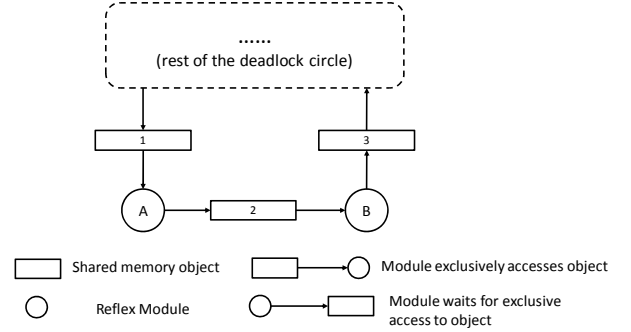

Figure 9: The Assumed Deadlock Situation. In this situation modules circularly wait for shared memory objects that are exclusively accessed by each other.

to object set 2. Recalling that the weakest module in a sharing group always owns the object, we have:

$$\text{Owner}(1) <= A, \text{ and Owner}(2) <= A$$

In Reflex, there are only three possible ways for A to reach such a situation.

In the first way, A does not own object set 1 or 2. A has got exclusive access to set 1 from Owner(1), and is requesting Owner(2) for exclusive access to object set 2. Thus, according to the second regulation (total order requesting) described in Section 5.5, we have

$$\text{Owner}(1) < \text{Owner}(2) < A \quad (1)$$

In the second way, A does not own object set 1 but does own 2. A has got exclusive access to object set 1 from Owner(1) and is stalling for a release request of object set 2. We have:

$$\text{Owner}(1) < \text{Owner}(2) = A \quad (2)$$

In the third way, A owns both object set 1 and 2. A is stalling for a release request of object set 2. We have

$$\text{Owner}(1) = A = \text{Owner}(2) \quad (3)$$

Combing (1)-(3), we have the following relations for any module in the deadlock circle and the two related owners:

$$\text{Owner}(1) = A = \text{Owner}(2) \quad (4)$$
$$\text{or, Owner}(1) < \text{Owner}(2) <= A \quad (5)$$

By applying (4) and (5) to each module in the deadlock circle, we have

$$\text{Owner}(1) <= \ldots \text{Owner}(n) <= \text{Owner}(1) \quad (6)$$

If (5) holds for any module in the circle, (6) is equivalent to Owner(1)<Owner(1), a contradiction. If (4) holds for all modules, we have Owner(1)=…=Owner(n)=A=B… This is again impossible, simply because a module cannot lock itself in Reflex DSM.

Therefore, the assumed general deadlock situation can never happen. Reflex DSM prevents deadlock.